# A Pd/Al$_2$O$_3$-based micro-reformer unit fully integrated in silicon technology for H-rich gas production

**M Bianchini**[1], **N Alayo**[1], **L Soler**[3], **M Salleras**[2], **L Fonseca**[2], **J Llorca**[3] and **A Tarancon**[1,4]

[1] Catalonia Institute for Energy Research (IREC), Department of Advanced Materials for Energy, 08930, Sant Adriá del Besòs, Barcelona, Spain
[2] IMB-CNM (CSIC), Institute of Microelectronics of Barcelona, National Center of Microelectronics, CSIC, Campus UAB, 08193, Bellaterra, Barcelona, Spain
[3] Institut de Tècniques Energètiques i Departament d'Enginyeria Química, Universitat Politècnica de Catalunya, EEBE, 08019 Barcelona, Spain
[4] ICREA, 08010, Barcelona, Spain

mbianchini@irec.cat

**Abstract**. This work reports the design, manufacturing and catalytic activity characterization of a micro-reformer for hydrogen-rich gas generation integrated in portable-solid oxide fuel cells (μ-SOFCs). The reformer has been designed as a silicon micro monolithic substrate compatible with the mainstream microelectronics fabrication technologies ensuring a cost-effective high reproducibility and reliability. Design and geometry of the system have been optimized comparing with the previous design, consisting in an array of more than 7x10$^3$ vertical through-silicon micro channels perfectly aligned (50 μm diameter) and a 5 W integrated serpentine heater consisting of three stacked metallic layers (TiW, W and Au) for perfect adhesion and passivation. Traditional fuels for SOFCs, such as ethanol or methanol, have been replaced by dimethyl ether (DME) and the chosen catalyst for DME conversion consists of Pd nanoparticles grafted on an alumina active support. The micro-channels have been coated by atomic layer deposition (ALD) with amorphous Al$_2$O$_3$ and the influence of rapid thermal processing (RTP) on such film has been studied. A customized ceramic 3D-printed holder has been designed to measure the specific hydrogen production rates, DME conversion and selectivity profiles of such catalyst at different temperatures.

## 1. Introduction

Due to the increasing energy demand for small-scale portable power sources and to the limited amount of specific energy provided by the currently employed devices, in recent years many researchers have focused on finding new technologies to fulfill this energy gap especially for applications requiring low power[1]. One of the best candidate for this purpose are micro-solid oxide fuel cells (μ-SOFCs) which combine fuel flexibility, high efficiency and energy density up to 5 times the one of Li-ion batteries[2]. The most common approach to miniaturize SOFCs is based on the monolithic integration of functional free-standing membranes in silicon technology[1-7], although many technological challenges still remain unsolved.

One of the most challenging issues in order to commercialize these devices is to safety ensuring a continuous hydrogen flow to be feed the fuel cells stack. The use of catalytic reformers for on-board hydrogen production is an established alternative to direct hydrogen storage, since hydrocarbon-based fuels provide high energy density, low cost, safety, and easy transportation[8].

Conversely to other micro-reactors reported in literature which are based on packed bed reactor[9] or in-plane micro-channels[10-12], our group demonstrated the feasibility of the fabrication of a silicon monolithic micro-reformer consisting of an array of vertical micro-channels (50 μm diameter) and a 5 W serpentine heater by means of mainstream microelectronics fabrication processes (photolithography, wet etching, physical and chemical vapor deposition and reactive ion etching), leading to fast start-up time (less than a minute) and low energy consumption below 500 J[13].

This micro-reactor has first been designed for ethanol steam reforming and methanol dry reforming by coating inside the micro-channels a $SiO_2/CeO_2$ active support film with Pd/Rh nanoparticles grafted on top[14]. A modification of the heterogeneous catalyst coated inside the vertical micro-channels allows to adapt the reactor to the conversion of a different fuel, e.g. dimethyl ether (DME). In fact, DME has been chosen as a non-petroleum based alternative fuel that could have a great impact on society due to its low environmental impact and physio-chemical properties that would lead to technological advantages compared to other fuels such as methane or ethanol[15].

According to previous studies from Semelsberger et al.[16] DME demonstrates higher well-to-wheel efficiency and lower greenhouse gases emissions together with the lowest start-up energies correlating to higher overall efficiencies for on-board automotive fuel processors. Finally, DME storage and transport infrastructures are the same as LPG or natural gas, making it cheaper to integrate than compared to hydrogen. On the other hand, from a technological point of view employing DME would solve the problems linked to the thermal management coming from high reforming temperature of ethanol (above 700°C), since the DME reforming reaction occurs at much lower temperatures[17].

A customized ceramic 3D-printed (3DCERAM SLA) holder has been designed to measure the specific hydrogen production rates, DME conversion and selectivity profiles of such catalyst upon increasing temperature for steam reforming reaction at steam-to-carbon ratio of 1.5 (1) and partial oxidation (2) which latter is the ideal conversion reaction for portable applications[18] although resulting in lower hydrogen yield:

$$CH_3OCH_3 + 3H_2O \rightarrow 6H_2 + 2CO_2 \quad (1)$$
$$CH_3OCH_3 + \frac{1}{2}O_2 \rightarrow 3H_2 + 2CO \quad (2)$$

## 2. Fabrication process

The micro-reformer design proposed by our group[14] was specifically thought to reduce the pressure drop and the energy consumption, to accelerate the system start-up and to thermally insulate the active zone by etching a trench into the bulk silicon. The fabrication process was finally optimized and it is schematically illustrated in figure 1.

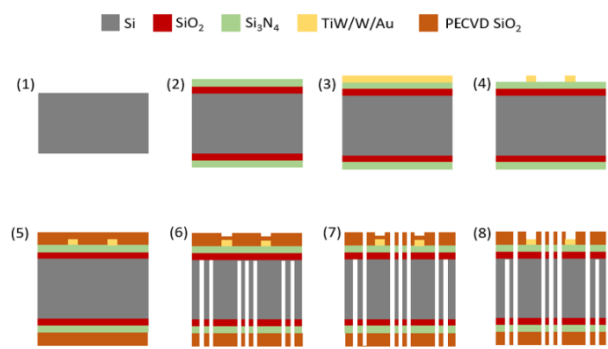

Figure 1. (1) 500 μm-thick single crystal silicon wafer of 4'' diameter. (2) 100 nm-thick thermal $SiO_2$ and 300 nm-thick $Si_3N_4$ by PECVD. (3) 330 nm-thick triple layer of TiW/W/Au by sputtering. (4) Heater defined by lift-off. (5) 2.5 μm-thick SiO2 deposited by PECVD. (6) DRIE of channels and trench through dielectrics and bulk silicon from the back side. (7) DRIE from the top side to open the channels. (8) HF wet etching to remove the $SiO_2$ on the top of the contacts.

A 500 μm-thick single crystal (100)-oriented p-type silicon double side polished wafer is thermally oxidized to grow 100 nm of $SiO_2$ on both sides, on top of which 300 nm of $Si_3N_4$ are deposited by plasma-enhanced chemical vapor deposition (PECVD). Subsequently, the heater is defined by sputtering a triple layer of 30 nm TiW (for good adhesion), 250 nm of W and 50 nm of Au (passivation) by lifting-off with acetone the cured photoresist below the sputtered layer.

The following step involves the deposition of 2.5 μm of $SiO_2$ on the top and bottom of the wafer by PECVD that will be used as a mask for the dry etching process. A deep reactive ion etching (DRIE) process follows the lithography that defines the trench and channels on the bottom of the wafer and it takes place in two steps: the first to etch the dielectric layer and the second to etch the bulky silicon. Just before repeating the same process on the top of the wafer, DRIE is employed on the top of the contacts to reduce the thickness of the oxide to 1.5 μm. The channels are finally opened by DRIE and the oxide left on the contacts is removed by wet etching in 5% diluted HF solution.

The second part of the fabrication process, i.e. the catalyst fabrication, consisted in the deposition of a 100 nm-thick $Al_2O_3$ support by atomic layer deposition (ALD) along the micro-channels walls by alternating pulses of the Al precursor trimethylaluminium (TMA) and $H_2O$ as oxidant. Finally, after a rapid thermal process (RTP) to crystallize the alumina layer was performed, a uniform dispersion of Pd nanoparticles on top of the support was obtained by infiltration of a $Pd(NO_3)_2$ acid solution (in $HNO_3$, pH=1-2) inside the micro-channels. It is noticeable that before infiltrating with the precursor solution, water was made pass through the active area in order to increase the number of hydroxyl groups on the surface of the alumina layer increasing the chance of palladium grafting. Finally, the reactor was dried overnight at 100ºC, calcinated at 500ºC for 3h and finally reduced in a 10% $H_2/N_2$ flow for 1h at 300ºC before starting the activity measurement.

## 3. Results

Figure 2 reports the result of the optimized microfabrication process described above. The so fabricated array of vertical microchannels ensures high surface area within which the alumina catalytic support was deposited, and a complete coverage of the walls with a conformal amorphous layer was obtained. RTP was performed and the influence of the temperature on the optical and structural properties is reported in figure 3 together with the SEM image of the 100 nm-thick alumina layer. Firstly, since the material is transparent in the spectral range of the ellipsometric measurements, its refractive index was modeled with the dispersion relation of Cauchy[19]. It is noticeable that the optical properties of alumina drastically change above 850ºC, when the refractive index at 532 nm wavelength change from 1.66 to 1.72 as expected after crystallization. Conflicting results obtained from the sample annealed at 800ºC suggest this is the temperature at which crystallization starts to occur.

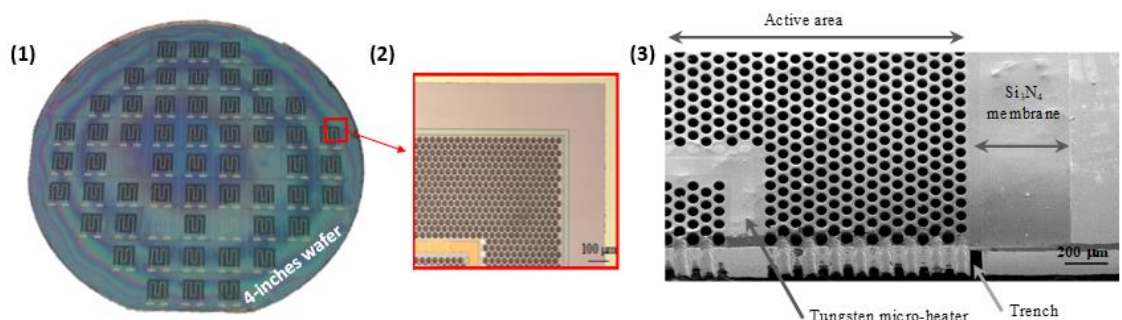

Figure 2. (1) Top view of the micromachined wafer. (2) Optical microscope image (10X) of the active area. (3) SEM cross-section image of the micro-reformer after the fabrication. Image (2) and (3) readapted from reference [14].

Secondly, grazing incident angle X-ray diffraction measurements (GIXRD) was performed to obtain diffraction spectra without the strong signals from the silicon single crystal substrate. The characteristic peaks showed in figure 3.4 belong to the gamma and theta phases of alumina which start crystallizing simultaneously. Moreover, it is noticeable that increasing the annealing temperature the crystallinity of the film increases and the peak shift around 2θ ~ 47º suggest an increase of the theta phase above 950ºC.

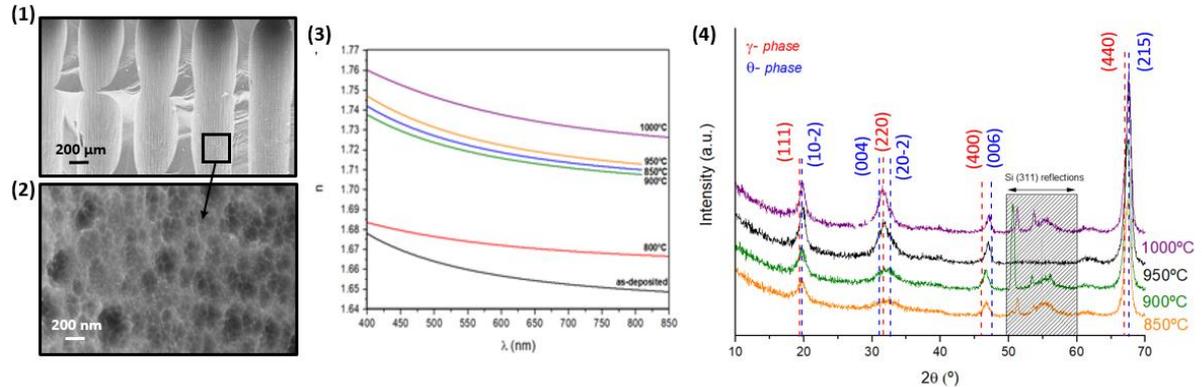

Figure 3. (1) Cross-section SEM image of the micro-channels. (2) SEM image of the 100 nm-thick as-deposited alumina layer. (3) Ellipsometric measurement in the visible range of the as-deposited $Al_2O_3$ film and after RTP. (4) GIXRD spectra of the annealed sample, measurements performed at an optimized angle of ω=0.3º.

Finally, a fairly uniform dispersion of palladium nanoparticles was obtained by infiltration and observed by means of a SEM backscattered electrons detector. The analysis was performed on 5 different channels and at various positions, demonstrating that palladium was present along the whole active area of the micro-reformer. The mean particles size is estimated to be 28 nm.

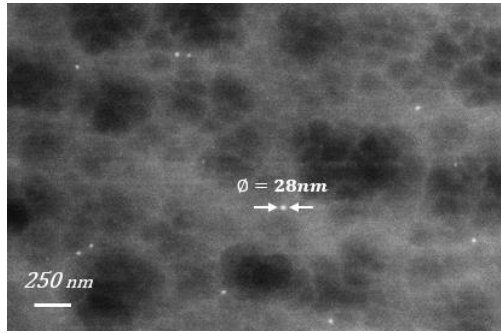

Figure 4. Cross-section Backscattered electrons SEM image representing the Pd nanoparticles dispersion by Z-contrast with the alumina support layer

Figure 5 shows the customized 3D-printed holder designed for the catalytic activity measurements in the temperature range between 400 and 650ºC. Before the experiment, the gas tightness of the setup has been tested with positive results, confirming the choice of the alumina felt sealing. Experiments are currently ongoing.

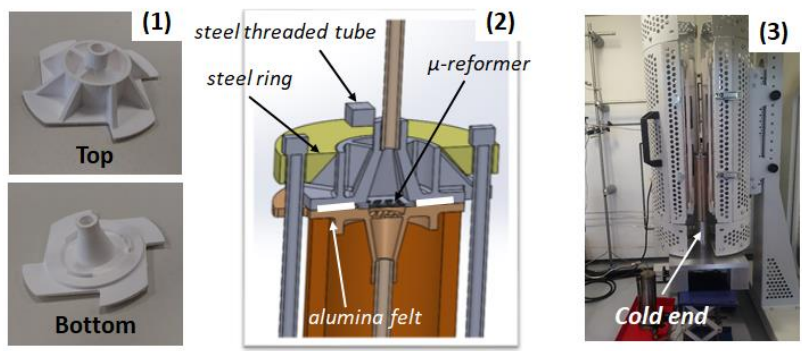

Figure 5. (1) Bottom and top view of the 3D-printed holder. (2) Cross-section view of the 3D-printed holder. (3) Picture of the setup mounted inside the furnace.

## 4. Conclusions

The wafer-level microfabrication process for the realization of a micro-reformer integrated in silicon technology has been successfully optimized. High surface area, ease of integration of the catalytic system and fuel flexibility are the main features of the fabricated reactor, besides requiring low energy for a rapid startup, that make it an ideal device for portable applications. The heterogeneous catalyst for DME reforming has been deposited within the active part of the device and it has been structurally characterized. Finally, the setup for the catalytic measurements has been design and 3D-printed, ensuring gas-tightness and thermomechanical stability up to 650ºC.

## 5. Acknowledgments

This project has received funding from the European Research Council (ERC) under the European Union's Horizon 2020 research and innovation programme (ERC-2015-CoG, grant agreement No #681146- ULTRA-SOFC). Special thanks go to Josep M. Bassas for GIXRD measurements and to the IMB-CNM clean-room staff.